\documentclass[fleqn,twoside]{article}
\usepackage[headings]{espcrc2}
\usepackage{pslatex}

\usepackage{graphicx}

\def\pt{{p_{\mbox{\scriptsize T}}}}
\def\ptcut{{p_{\mbox{\scriptsize T}}^{\mbox{\scriptsize cut}}}}

\def\mathswitchr#1{\relax\ifmmode{\mathrm{#1}}\else$\mathrm{#1}$\fi}
\newcommand{\PH}{\mathswitchr H}
\newcommand{\Pp}{\mathswitchr p}
\newcommand{\Pt}{\mathswitchr t}

\newcommand{\GeV}{\mbox{GeV}}
\newcommand{\MeV}{\unskip\,\mathrm{MeV}}

\def\mathswitch#1{\relax\ifmmode#1\else$#1$\fi}
\newcommand{\Mt}{\mathswitch {m_\Pt}}

\def\Tab#1{{Tab.~\ref{#1}}}
\def\Eq#1{{Eq.~(\ref{#1})}}
\def\Ref#1{{Ref.~\cite{#1}}}
\def\Refs#1{{Refs.~\cite{#1}}}
\def\Fig#1{{Fig.~\ref{#1}}}

\def\onejet{{\mbox{1-jet}}}
\title{Top-quark pair + 1-jet production at next-to-leading order QCD
  \begin{picture}(0,0)
    \put(-10.,100.){\parbox{4cm}{
        \begin{flushright}
          \small MPP-2008-75 \\ 
          MZ-TH/08-21\\ 
          SFB/CPP-08-45          
        \end{flushright}
        }}
  \end{picture}}

\author{%
  S. Dittmaier\address{Max-Planck-Institut f\"ur Physik
    (Werner-Heisenberg-Institut), D-80805 M\"unchen, Germany}%
  \thanks{Supported
    in part by the European Community's Marie-Curie Research 
    Training Network HEPTOOLS under contract
    MRTN-CT-2006-035505},
  P. Uwer\address{Institut f\"ur Theoretische Teilchenphysik, 
    Universit\"at Karlsruhe, D-76128 Karlsruhe, Germany}%
  \thanks{Supported as Heisenberg Fellow of the 
    Deutsche Forschungsgemeinschaft DFG and by the DFG 
    Sonderforschungsbereich/Transregio 9 
    "Computergest\"utzte Theoretische Teilchenphysik" SFB/TR9.%
    },
  S. Weinzierl\address{Universit\"at Mainz, D-55099 Mainz, Germany}
  }

\runtitle{Top-quark pair + 1-jet production at next-to-leading order QCD}
\runauthor{P.Uwer}

\begin{document}
\begin{abstract}
Top-quark pair production with an additional jet is an important
signal and background process at the LHC. We present the 
next-to-leading order QCD calculation for this process and show
results for integrated as well as 
differential cross sections.
\end{abstract}

\maketitle

\section{Introduction}
With a mass of 172.6 GeV the top-quark is by far the heaviest
elementary fermion in the Standard Model (SM). Its mass is more than 30 times larger
than the mass of the next heaviest fermion, the bottom-quark. 
The large mass has lead to
various speculations whether the top-quark behaves as a normal quark
or whether it plays a special role.
In particular, the fact that the top-quark mass is close to the scale of
electroweak symmety breaking---or equivalently, that the Yukawa coupling to
the Higgs is very close to one---has motivated different scenarios
in which the top-quark drives the electroweak symmetry breaking. 
For a recent overview we refer the interested reader to 
\Refs{Bernreuther:2008ju,Han:2008xb}.
In the context of the SM the top-quark interactions are
completely determined through the gauge structure.
The only free parameter appearing in top-quark
physics is the top-quark mass. Once this parameter is measured all
remaining properties are predicted. An important task for
the ongoing Tevatron collider and the upcoming LHC is the precise
measurement of the top-quark properties. In this context the
production of a top-quark pair togehter with an additional jet is
an important reaction. This becomes already clear from the simple
observation that a substantial number of events in the inclusive 
top-quark sample is accompanied by an additional jet. Depending on the
energy of the additional jet the fraction of events with an additional
jet can easily be of the order of 10--30\%.
For a more precise understanding of the topology of top-quark events
it is thus important to have also an improved understanding of 
top-quark pair production together with a jet. 
Recently it has been argued that the inclusive top-quark pair cross
section can be measured at the LHC with an experimental uncertainty of
about 5\%. The available theoretical predictions at next-to-leading
order (NLO)  togehter with
resummation lead to a theoretical uncertainty of more than 
10\% \cite{Moch:2008qy,Cacciari:2008zb,Kidonakis:2008mu}.
The dominant uncertainy comes from the residual scale dependence, while
the uncertainties  due to the parton distribution functions are
comparably small. 
In \Ref{Moch:2008qy} parts of the full next-to-next-to-leading order
(NNLO) were derived from general
arguments. It has been shown that an NNLO calculation will reduce
the theoretical uncertainty down to 5\%. Apart from the two-loop
corrections (where progress has been made recently 
\cite{Czakon:2007ej,
Czakon:2008zk}) 
one important ingredient in going to NNLO are the one-loop
corrections to $\Pt\bar \Pt + \onejet$ production. 
Apart from its
significance as signal process it turns out that $\Pt \bar \Pt +
\onejet$ production is also an important background to 
various new physics searches. A prominent example is Higgs production
via vector-boson fusion. This reaction represents an important
discovery channel for a SM Higgs boson in the mass range
of 120--180~GeV. The major background to this reaction is due to 
$\Pt \bar \Pt + \onejet $ \cite{Alves:2003vp}, again underlining
the need for precise theoretical predictions for this process.
It is well known that predictions at leading-order (LO) in the 
coupling constant of QCD are plagued 
by large uncertainties. In many
cases the LO predictions in QCD give 
only a rough estimate. Only by including NLO corrections
a quantitative reliable prediction can be obtained. Given that the
conceptual problems of doing such calculations are solved since quite
some time, one might think that doing the required calculations should
be a straightforward task. Unfortunately it turns out that this is not
the case. The calculation of QCD corrections for $2 \to 3$ and  $2 \to 4$
reactions is still a highly non-trivial task---not speaking about
reactions with an even higher multiplicity. 
In general the problem can be
attributed to the fact that the corrsponding matrix elements are
complex
functions of many variables so that an analytic treatment is no longer
feasible due to the large size of the expressions. A solution to this
problem is to resort to numerical methods. While in principle fine
one is in many cases plagued by numerical instabilities and the long runtime. 
In particular, the reduction of one-loop tensor integrals to scalar
one-loop integrals is in general difficult to do in a numerically stable
way. In that context the calculation of the one-loop corrections
to top-quark pair production with an additional jet is also
interesting as a benchmark process for the developement of new
methods. 
In the present article we will briefly comment on the calculation
of the NLO corrections. 
In addition we
will show results for integrated quantities as well as for differential 
distributions.

\section{Calculation}
\subsection{Born approximation}
In Born approximation the partonic reactions are 
$  gg\to \Pt\bar \Pt g$,
$  q\bar q \to \Pt\bar \Pt g$,
$  qg \to \Pt\bar \Pt q$, and
$  g\bar q \to \Pt\bar \Pt \bar q$.
The last three reactions are related by
crossing. In Born approximation various well-tested approaches to
calculate the required matrix elements exist. We used recurrence
relation \`a la Berends and Giele \cite{Berends:1987me} 
and a Feynman-diagram-based
approach. In both cases four-dimensional helicity methods were
employed. In addition we also used {\sl Madgraph}~\cite{Stelzer:1994ta} for
checking. We found complete 
agreement of the different methods. The explicit LO calculation
shows indeed the large scale dependence as expected. 
Without going into details we just mention that in LO the
importance of the individual partonic channels does not follow the
pattern known from inclusive top-quark pair production. While at the
Tevatron the situation $\Pt\bar \Pt + \onejet $ production is similar to
the inclusive reaction, that is the total cross section is dominated
by the quark--anti-quark channel followed by the gluon-fusion 
process, the situation at the LHC is different from inclusive 
production. The most important channel is gluon fusion, but in contrast
to inclusive production the second important channel is the
$qg$-channel. This is due to the large parton luminosity for this
channel and due to a sizeable partonic cross section.
  
\subsection{Virtual corrections}
The virtual corrections consist of the one-loop corrections to the
LO reactions. One can classify the
corrections into self-energy corrections, vertex corrections, box-type 
corrections, and pentagon-type corrections where all the external legs
are connected to
one loop thus forming a pentagon. The latter are the most complicated
ones owing to their complexity and the involved tensor integrals. The
challenging step in this context is the numerically fast and stable
reduction of the tensor integrals to scalar one-loop integrals. 
To ensure the correctness of our results we did two independent
calculations of the virtual corrections using as far as possible
different methods and also different tools.
In one calculation the virtual corrections are essentially obtained 
following the procedure described in \Ref{Beenakker:2002nc}, where
$\Pt\bar \Pt + \PH$ production at hadron colliders was considered.
Feynman diagrams and amplitudes have been generated with the
{\sl FeynArts} package \cite{Kublbeck:1990xc,Hahn:2000kx}
and further processed with in-house {\sl Mathematica} routines,
which automatically create an output in {\sl Fortran}.
The IR (soft and collinear) singularities are analytically separated
from the finite remainder as described in
\Refs{Beenakker:2002nc,Dittmaier:2003bc}.
The tensor integrals appearing in the pentagon diagrams
are directly reduced to box 
integrals following \Ref{Denner:2002ii}. This method does not
introduce inverse Gram determinants in this step, thereby avoiding
notorious numerical instabilities in regions where these determinants
become small. Box and lower-point integrals are reduced 
\`a la Passarino--Veltman \cite{Passarino:1978jh} to scalar integrals,
which are either calculated analytically or using the results of
\Refs{'tHooft:1978xw,Beenakker:1988jr,Denner:1991qq}. 
Sufficient numerical stability is already achieved in this
way. Nevertheless the integral evaluation is currently further refined
by employing the more sophisticated methods described in
\Ref{Denner:2005nn} in order to numerically stabilize the tensor
integrals in exceptional phase-space regions.

In the second calculation the evaluation of loop diagrams starts
with the generation of diagrams and amplitudes via {\sl QGRAF} 
\cite{Nogueira:1991ex},
which are then further manipulated with {\sl Form}
\cite{Vermaseren:2000nd} and automatically translated into {\sl C++} code.
The reduction of the 5-point tensor integrals to scalar
integrals is performed with an extension of the method described in 
\Ref{Giele:2004iy}. In this procedure also
inverse Gram determinants of four four-momenta are avoided.
The lower-point tensor integrals are reduced
using an independent implementation of the Passarino--Veltman procedure.
The IR-finite scalar integrals are
evaluated using the {\sl FF} package 
\cite{vanOldenborgh:1990wn,vanOldenborgh:1991yc}.

\subsection{Real corrections}
The matrix elements for the real corrections are given by
$0 \to \Pt \bar \Pt g g g g$, 
$0 \to \Pt \bar \Pt q \bar q g g$,
$0 \to \Pt \bar \Pt q \bar q q' \bar q'$
and
$0 \to \Pt \bar \Pt q \bar q q \bar q$.
The various partonic processes are obtained from these matrix elements 
by all possible crossings of light particles into the initial state.

The evaluation of the real-emission amplitudes is
again performed in two independent ways.
Both evaluations employ 
the dipole subtraction formalism 
\cite{Catani:1996vz,Phaf:2001gc,Catani:2002hc}
for the extraction of IR singularities and for their
combination with the virtual corrections. 

One calculation of the real corrections 
results from a fully automated calculation
based on helicity amplitudes, as
described in \Ref{Weinzierl:2005dd}.
Individual helicity amplitudes are computed with the help of
Berends--Giele recurrence relations \cite{Berends:1987me}.
The evaluation of color factors and the generation of subtraction
terms is automated.
For the channel $g g \to \Pt \bar \Pt g g$ a dedicated
soft-insertion routine \cite{Weinzierl:1999yf} is used for the generation 
of the phase space.
The second calculation uses for the LO $2 \to 3$ processes
and the $g g \to \Pt \bar \Pt g g$ process optimized code obtained from
a Feynman diagrammatic approach.  As in the calculation described
before standard techniques
like color decomposition and the use of helicity amplitudes are
employed. For the $2 \to 4$ processes including light quarks, {\sl
  Madgraph} \cite{Stelzer:1994ta} has been used. The subtraction terms
according to \Ref{Catani:2002hc} are obtained in a semi-automatized
manner based on an in-house library written in {\sl C++}.

\section{Results}

In the following we consistently use the CTEQ6 
\cite{Pumplin:2002vw,Stump:2003yu}
set of parton distribution functions (PDFs). In detail, we take
CTEQ6L1 PDFs with a 1-loop running $\alpha_{\mathrm{s}}$ in
LO and CTEQ6M PDFs with a 2-loop running $\alpha_{\mathrm{s}}$
in NLO.
The number of active flavours is $N_{\mathrm{F}}=5$, and the
respective QCD parameters are $\Lambda_5^{\mathrm{LO}}=165\MeV$
and $\Lambda_5^{\overline{\mathrm{MS}}}=226\MeV$.
Note that the top-quark loop in the gluon self-energy is
subtracted at zero momentum. In this scheme the running of 
$\alpha_{\mathrm{s}}$ is generated solely by the contributions of the
light quark and gluon loops. The top-quark mass is
renormalized in the on-shell scheme, as numerical value we take $\Mt=174\GeV$.

We apply the jet algorithm of \Ref{Ellis:1993tq}
with $R=1$ for the definition of the tagged hard jet. Unless stated otherwise
we require a transverse momentum of
$p_{\mathrm{T,jet}}>\ptcut = 20\GeV$ for 
the hardest jet.
The outgoing (anti-)top-quarks are neither affected
by the jet algorithm nor by the phase-space cut.
Note that the LO prediction and the virtual corrections are not influenced
by the jet algorithm, but the real corrections are.
\looseness-1

\begin{figure*}
  \centerline{
    \leavevmode
    \includegraphics[width=0.96\columnwidth]{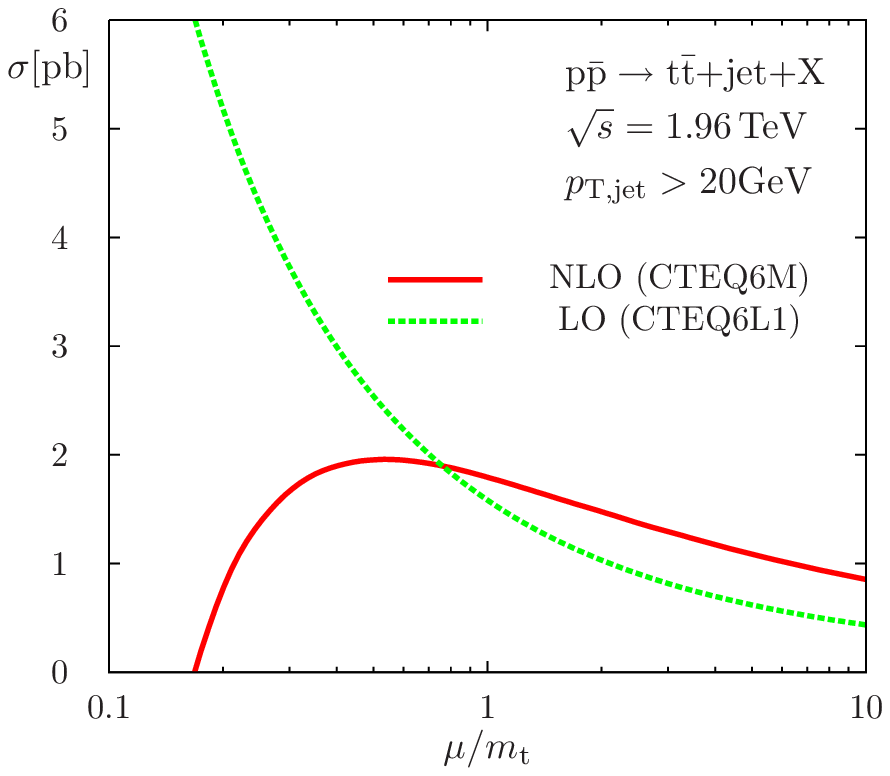}
    \hspace*{2em}
    \includegraphics[width=0.95\columnwidth]{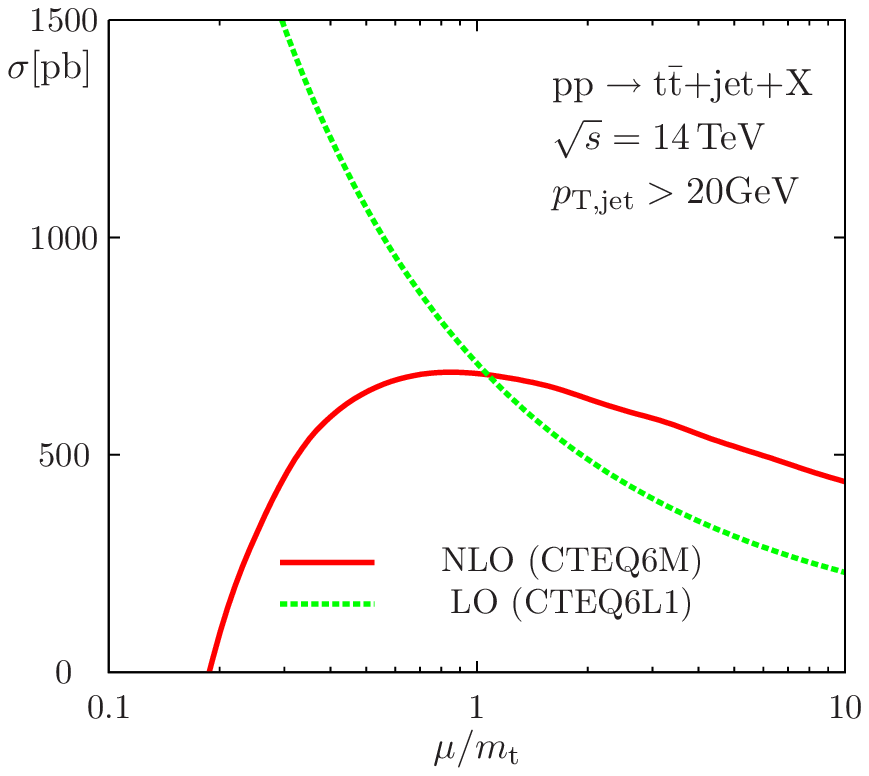}
  } \vspace*{-2.6em}
    \caption{Scale dependence of the LO and NLO cross sections for
      $\Pt\bar \Pt + \onejet$ production at the Tevatron (left) and the LHC 
      (right) as taken from \Ref{Dittmaier:2007wz}, 
      with the renormalization scale ($\mu_r$) and the factorization scale
      ($\mu_f$) set to~$\mu$.}
    \label{fig:NLOcs}
\end{figure*}
In \Fig{fig:NLOcs} 
the scale dependence of the NLO cross sections
is shown. For comparison, the LO results are included as well. Varying
the scale in the usual range that is a factor 2 up and down around a
central scale we observe a reduction of the scale dependence of a
factor 3 at Tevatron and about a factor 6 at the LHC. From the absolute 
values of the cross sections one concludes that indeed a significant 
contribution to the inclusive top-quark pair cross section comes from 
$\Pt \bar \Pt + \onejet$. For reasonable scale choices one can also
observe that the NLO corrections are of moderate size.
In particular around $\mu=\Mt$ the corrections are small. 
The interference of $C$-odd contributions of the amplitude with
$C$-even parts, where $C$ denotes the charge conjugation, produces 
a forward--backward charge asymmetry at the Tevatron. For the inclusive
top-quark pair sample, where this asymmetry is an NLO effect, 
this has been studied in  
\Refs{Halzen:1987xd,Kuhn:1998kw,Bowen:2005ap}. In the case
of $\Pt \bar \Pt + \onejet$ this asymmetry appears already in LO.
In LO the asymmetry is defined by
\begin{equation}
\label{eq:asym}
 A^{\Pt}_{\mathrm{FB,LO}} = 
\frac{\sigma^-_{\mathrm{LO}}}{\sigma^+_{\mathrm{LO}}},
\quad
\sigma^\pm_{\mathrm{LO}} = 
\sigma_{\mathrm{LO}}(y_{\Pt}{>}0)\pm\sigma_{\mathrm{LO}}(y_{\Pt}{<}0),
\end{equation}
where $y_{\Pt}$ denotes the rapidity of the top-quark.
Cross-section contributions 
$\sigma(y_{\Pt}$ \raisebox{-.2em}{$\stackrel{>}{\mbox{\scriptsize$<$}}$} $0)$ 
correspond to top-quarks in the forward or backward hemispheres, respectively,
where incoming protons fly into the forward direction by definition.
Denoting the corresponding NLO contributions to the cross sections by
$\delta\sigma^\pm_{\mathrm{NLO}}$,
we define the asymmetry at NLO by
\begin{equation}
\label{eq:NLOasym}
 A^{\Pt}_{\mathrm{FB,NLO}} = 
\frac{\sigma^-_{\mathrm{LO}}}{\sigma^+_{\mathrm{LO}}}
\left( 1+
 \frac{\delta\sigma^-_{\mathrm{NLO}}}{\sigma^-_{\mathrm{LO}}}
-\frac{\delta\sigma^+_{\mathrm{NLO}}}{\sigma^+_{\mathrm{LO}}} \right),
\end{equation}
i.e.\ via a consistent expansion in $\alpha_{\mathrm{s}}$.
Note, however, that the LO cross sections in Eq.~(\ref{eq:NLOasym})
are evaluated in the NLO setup (PDFs, $\alpha_{\mathrm{s}}$).
\begin{figure}
  \centerline{  
    \leavevmode
    \includegraphics[width=\columnwidth]{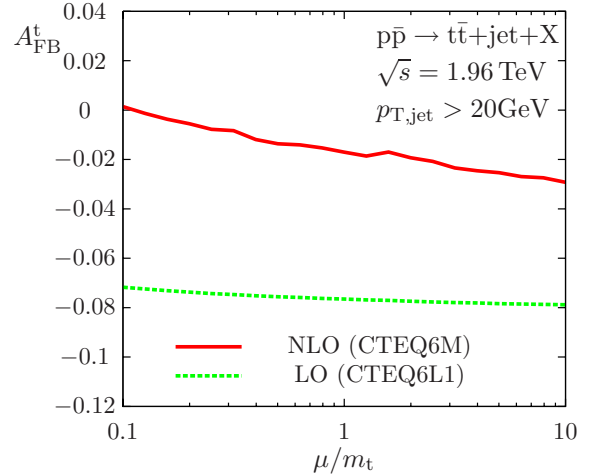}
  } \vspace*{-2.6em}
    \caption{Scale dependence of the LO and NLO forward--backward 
      charge asymmetry
      of the top-quark in $\Pp\bar\Pp\to\Pt\bar\Pt{+}$jet$+X$ at the
      Tevatron  as taken from \Ref{Dittmaier:2007wz} with 
      $\mu=\mu_f=\mu_r$.}
    \label{fig:NLOasym}
\end{figure}
The results for the asymmetry for different scale choices are shown
in \Fig{fig:NLOasym}. At LO we find an asymmetry of about
$-8$\%. The scale dependence is rather small. This is a consequence of
the fact that
$\alpha_s$ cancels exactly between the numerator and the
denominator. In addition the residual factorization scale dependence
also cancels to a large extent in the ratio. At NLO we find a large correction
compared to the LO result. The asymmetry is almost washed out at NLO. 
The scale dependence is increased in NLO which
seems natural given the small dependence in LO. To investigate the
origin of the large NLO corrections to the asymmetry we studied the 
dependence  on $\ptcut$, the minimal $\pt$ used to resolve the additional
jet.
\begin{table*}
  \caption{Cross section and forward-backward charge asymmetry at the
    Tevatron for
    different values of $\ptcut$ used to define the minimal
    transverse momentum $\pt$ of the additional jet
    ($\mu=\mu_f=\mu_r = m_t$).
    The upper and lower indices are the shifts towards $\mu = m_t/2$
    and $\mu = 2 m_t$.}
  \label{tab:XSection}
  \renewcommand{\tabcolsep}{1.5pc} 
  \renewcommand{\arraystretch}{1.2} 
  \begin{tabular}{c|l|l|l|l}
    \hline
    &\multicolumn{2}{|c|}{cross section [pb]}
    &\multicolumn{2}{|c}{ charge asymmetry [\%]}\\ 
    $\ptcut$ [GeV]&\multicolumn{1}{|c|}{LO}&
    \multicolumn{1}{|c|}{NLO}&
    \multicolumn{1}{|c|}{LO}
    &\multicolumn{1}{|c}{NLO}\\ \hline 
    20& 1.583(2)$^{+0.96}_{-0.55}$ & 1.791(1)$^{+0.16}_{-0.31}$ &
    $-7.69(4)^{+0.10}_{-0.085}$ & $-1.77(5)^{+0.58}_{-0.30}$ \\
    30& 0.984(1)$^{+0.60}_{-0.34}$ & 1.1194(8)$^{+0.11}_{-0.20}$ &
    $-8.29(5)^{+0.12}_{-0.085}$ & $-2.27(4)^{+0.31}_{-0.51}$ \\
    40& 0.6632(8)$^{+0.41}_{-0.23}$ & 0.7504(5)$^{+0.072}_{-0.14}$ &
    $-8.72(5)^{+0.13}_{-0.10}$ & $-2.73(4)^{+0.35}_{-0.49}$ \\
    50& 0.4670(6)$^{+0.29}_{-0.17}$ & 0.5244(4)$^{+0.049}_{-0.096}$
    & $-8.96(5)^{+0.14}_{-0.11}$ & $-3.05(4)^{+0.49}_{-0.39}$\\
    \hline 
  \end{tabular}
\end{table*}
The results are shown in \Tab{tab:XSection}. A strong dependence of
the cross section on $\ptcut$ is observed. For all $\ptcut$ values we
find that the NLO corrections to the
cross section  are of moderate size. While the cross section evidently
has a large sensitivity on $\ptcut$, the dependence of the asymmetry
on $\ptcut$ is slightly less pronounced. On the other hand we find, 
independent of the chosen $\ptcut$,
a significant difference between LO and NLO 
for the asymmetry
indicating that the definition of the asymmetry  \Eq{eq:NLOasym} 
is not stable with respect to higher-order corrections independent of
the $\ptcut$ value.
\begin{figure}
  \centerline{  
    \leavevmode
    \includegraphics[width=\columnwidth]{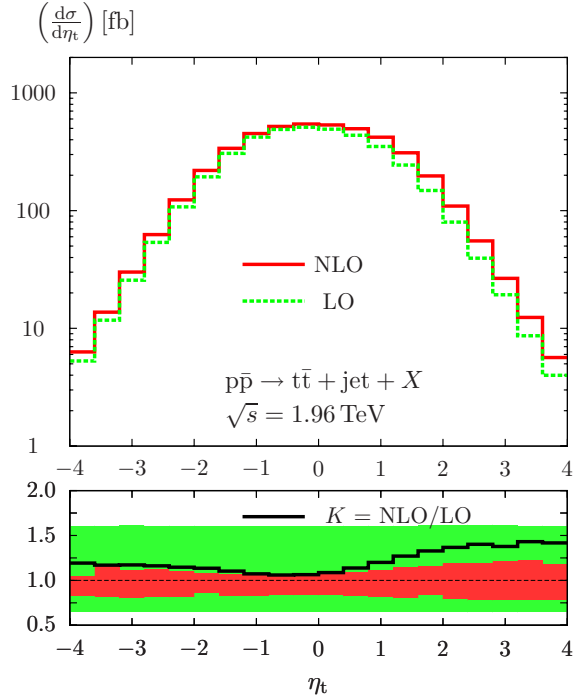}
  } \vspace*{-2.6em}
    \caption{Rapidity distribution of the top-quark. In the upper plot
    the thick (thin) line shows the NLO (LO) result. In the lower
    plot the thick line shows the $K$-factor, the two bands show the scale
    dependence in LO (outer band) and NLO (inner band) 
    when the scale ($\mu=\mu_r=\mu_f$) is
    varied by factor 2 up and down around the central value $\mu=\Mt$.}
    \label{fig:eta_top}
\end{figure}
\begin{figure}
  \centerline{
    \leavevmode
    \includegraphics[width=0.96\columnwidth]{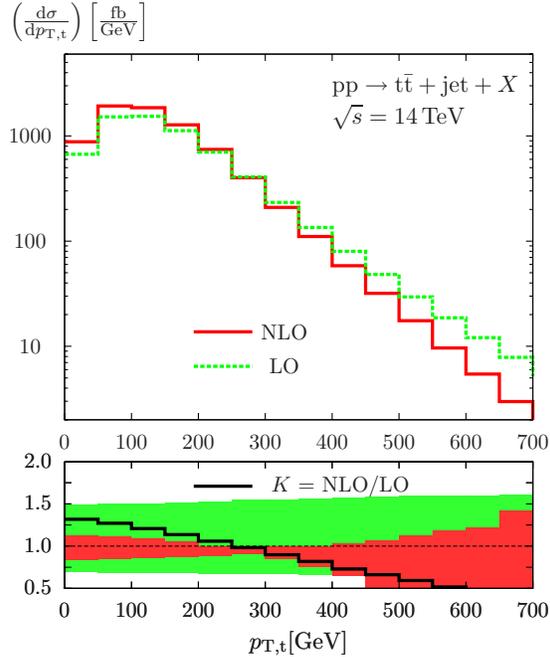}
  } \vspace*{-2.6em}
    \caption{Same as \Fig{fig:eta_top} for the $\pt$-distribution of
    the top-quark at LHC.}
    \label{fig:pt_top}
\end{figure}
In \Fig{fig:eta_top} the rapidity distribution for the top-quark is
shown. One can easily observe the asymmetry of the LO distribution leading
to the aforementioned $-8$\% for the integrated asymmetry. The NLO
corrections are different in shape compared to LO. They are larger in
the forward direction than in the backward direction. The result is
that the asymmetry is washed out by the NLO correction.
In the lower plot of \Fig{fig:eta_top} the $K$-factor and the scale
dependence is shown. The corrections are between $+$20 and $+$40\%.
The large scale dependence of the LO result is again significantly
improved when the NLO corrections are taken into account.
In \Fig{fig:pt_top} the $\pt$-distribution of the top-quark is shown---now for
the LHC. For not too large $\pt$ the corrections are of moderate size.
Starting with $+$30 \% for small $\pt$ and reaching $-$50 \% at 
$\pt\approx 450$ \GeV, indicating that the $K$-factor is strongly
phase-space dependent.  
Up to a $\pt$ of about 450 \GeV we also find an important improvement
of the scale uncertainty.

{\bf Conclusions:}
Predictions for $\Pt\bar\Pt{+}$jet production at hadron
colliders have been reviewed at NLO QCD. 
For the cross section the NLO corrections
drastically reduce the scale dependence of the LO predictions, which 
is of the order of 100\%.
The charge asymmetry of the top-quarks, which is measured at the
Tevatron, is significantly decreased at NLO and is almost washed out
by the residual scale dependence. In addition we have studied 
the $\ptcut$-dependence of the asymmetry in NLO. Further refinements
of the precise definition of the charge asymmetry are required
to stabilize the predictions with respect to higher-order corrections.
First results for differential distributions have been presented. 
The corrections are well under control
over a large phase-space region and the scale uncertainty is again 
improved compared to the LO results.


\begin{thebibliography}{10}

\bibitem{Bernreuther:2008ju}
W. Bernreuther,
\newblock J. Phys. G35 (2008) 083001. 
\newblock 

\bibitem{Han:2008xb}
T. Han,
\newblock (2008), arXiv 0804.3178.
\newblock 

\bibitem{Moch:2008qy}
S. Moch and P. Uwer,
\newblock (2008), arXiv 0804.1476.
\newblock 

\bibitem{Cacciari:2008zb}
M. Cacciari et~al.,
\newblock (2008), arXiv 0804.2800.
\newblock 

\bibitem{Kidonakis:2008mu}
N. Kidonakis and R. Vogt,
\newblock (2008), arXiv 0805.3844.
\newblock 

\bibitem{Czakon:2007ej}
M. Czakon, A. Mitov and S. Moch,
\newblock Phys. Lett. B651 (2007) 147, 
arXiv:0707.4139 
\newblock 


\bibitem{Czakon:2008zk}
M. Czakon,
\newblock (2008), arXiv:0803.1400.
\newblock 

\bibitem{Alves:2003vp}
A. Alves et~al.,
\newblock Phys. Rev. D69 (2004) 075005. 
\newblock 

\bibitem{Berends:1987me}
F.A. Berends and W.T. Giele,
\newblock Nucl. Phys. B306 (1988) 759.
\newblock 

\bibitem{Stelzer:1994ta}
T. Stelzer and W.F. Long,
\newblock Comput. Phys. Commun. 81 (1994) 357. 
\newblock 

\bibitem{Beenakker:2002nc}
W. Beenakker et~al.,
\newblock Nucl. Phys. B653 (2003) 151. 
\newblock 

\bibitem{Kublbeck:1990xc}
J. K{\"u}blbeck, M. B{\"o}hm and A. Denner,
\newblock Comput. Phys. Commun. 60 (1990) 165.
\newblock 

\bibitem{Hahn:2000kx}
T. Hahn,
\newblock Comput. Phys. Commun. 140 (2001) 418. 
\newblock 

\bibitem{Dittmaier:2003bc}
S. Dittmaier,
\newblock Nucl. Phys. B675 (2003) 447. 
\newblock 

\bibitem{Denner:2002ii}
A. Denner and S. Dittmaier,
\newblock Nucl. Phys. B658 (2003) 175. 
\newblock 

\bibitem{Passarino:1978jh}
G. Passarino and M.J.G. Veltman,
\newblock Nucl. Phys. B160 (1979) 151.
\newblock 

\bibitem{'tHooft:1978xw}
G. 't~Hooft and M.J.G. Veltman,
\newblock Nucl. Phys. B153 (1979) 365.
\newblock 

\bibitem{Beenakker:1988jr}
W. Beenakker and A. Denner,
\newblock Nucl. Phys. B338 (1990) 349.
\newblock 

\bibitem{Denner:1991qq}
A. Denner, U. Nierste and R. Scharf,
\newblock Nucl. Phys. B367 (1991) 637.
\newblock 

\bibitem{Denner:2005nn}
A. Denner and S. Dittmaier,
\newblock Nucl. Phys. B734 (2006) 62. 
\newblock 

\bibitem{Nogueira:1991ex}
P. Nogueira,
\newblock J. Comput. Phys. 105 (1993) 279.
\newblock 

\bibitem{Vermaseren:2000nd}
J.A.M. Vermaseren,
\newblock (2000), math-ph/0010025.
\newblock 

\bibitem{Giele:2004iy}
W.T. Giele and E.W.N. Glover,
\newblock JHEP 04 (2004) 029. 
\newblock 

\bibitem{vanOldenborgh:1990wn}
G.J. van Oldenborgh and J.A.M. Vermaseren,
\newblock Z. Phys. C46 (1990) 425.
\newblock 

\bibitem{vanOldenborgh:1991yc}
G.J. van Oldenborgh,
\newblock Comput. Phys. Commun. 66 (1991) 1.
\newblock 

\bibitem{Catani:1996vz}
S. Catani and M.H. Seymour,
\newblock Nucl. Phys. B485 (1997) 291. 
\newblock 

\bibitem{Phaf:2001gc}
L. Phaf and S. Weinzierl,
\newblock JHEP 04 (2001) 006. 
\newblock 

\bibitem{Catani:2002hc}
S. Catani et~al.,
\newblock Nucl. Phys. B627 (2002) 189. 
\newblock 

\bibitem{Weinzierl:2005dd}
S. Weinzierl,
\newblock Eur. Phys. J. C45 (2006) 745. 
\newblock 

\bibitem{Weinzierl:1999yf}
S. Weinzierl and D.A. Kosower,
\newblock Phys. Rev. D60 (1999) 054028. 
\newblock 

\bibitem{Pumplin:2002vw}
J. Pumplin et~al.,
\newblock JHEP 07 (2002) 012. 
\newblock 

\bibitem{Stump:2003yu}
D. Stump et~al.,
\newblock JHEP 10 (2003) 046. 
\newblock 

\bibitem{Ellis:1993tq}
S.D. Ellis and D.E. Soper,
\newblock Phys. Rev. D48 (1993) 3160. 
\newblock 

\bibitem{Dittmaier:2007wz}
S. Dittmaier, P. Uwer and S. Weinzierl,
\newblock Phys. Rev. Lett. 98 (2007) 262002. 
\newblock 

\bibitem{Halzen:1987xd}
F. Halzen, P. Hoyer and C.S. Kim,
\newblock Phys. Lett. B195 (1987) 74.
\newblock 

\bibitem{Kuhn:1998kw}
J.H. K{\"u}hn and G. Rodrigo,
\newblock Phys. Rev. D59 (1999) 054017, 
Phys. Rev. Lett. 81 (1998) 49. 
\newblock 

\bibitem{Bowen:2005ap}
M.T. Bowen, S.D. Ellis and D. Rainwater,
\newblock Phys. Rev. D73 (2006) 014008. 
\newblock 

\end{thebibliography}

\end{document}